# A MULTI RECONSTRUCTION STUDY OF BREAST DENSITY ESTIMATION USING DEEPLEARNING

*Vikash Gupta\*, Mutlu Demirer\*, Robert W. Maxwell, Richard D. White, Barbaros S. Erdal*

\* Vikash Gupta and Mutlu Demirer are joint first authors with equal contributions.

Center for Augmented Intelligence in Imaging,
Department of Radiology, Mayo Clinic,
Florida

## ABSTRACT

Breast density estimation is one of the key measurements in recognizing individuals predisposed to breast cancer. However, measuring breast density is often challenging because of low mammogram contrast and fluctuations in fatty tissue background. Most of the time, the breast density is estimated manually whereby a radiologist assigns one of the four density categories outlined in the Breast Imaging and Reporting Data System (BI-RADS). To increase the accuracy of breast density measurement beyond the quartile BI-RADS schema, efforts have been made at automating breast density classification.

Breast density estimation is one of the required tasks performed during interpretation of a mammogram. Dense breasts are more susceptible to breast cancer. The density estimation is challenging because of low contrast and fluctuations in mammograms' fatty tissue background. Traditional 2D mammograms are being replaced by tomosynthesis and its reconstructed variants (for example Hologic's Intelligent 2D and C-View). In an effort to decrease radiation dose, more screening centers are favoring the Intelligent 2D view and C-View. Deep-learning studies for breast density estimation use only a single acquisition for training a neural network. However, doing so restricts the number of images in the dataset. In this paper, we show that a neural network trained on all the acquisitions at once performs better than a neural network trained on any single acquisition. We discuss these results using the area under the receiver operator characteristic curves.

## 1. INTRODUCTION

Breast cancer is the most common form of identifiable and treatable cancer among women. Early diagnosis is the key to successfully treating such cancer. All known factors (gender, age, gene mutations, and family history) being constant, the amount of radiodense tissue is directly proportional to the risk of breast cancer.

Breast density is divided into four categories A to D in the order of increasing density. These categories are standardized and defined by Breast Imaging and Reporting and Data System (BI-RADS) [1]. The BI-RADS mammographic breast density criteria include four named categories: almost entirely fat, scattered fibroglandular density, heterogeneously dense, and extremely dense [2].

Density-based mammogram classification has always been subjective to the radiologist's interpretation. This subjective assessment leads to known significant intra- and interobserver variability and lack of gold-standard data. Most machine learning models are only as good as the training data provided. Before the proliferation of deep learning based models, several quantitative methods were used to measure breast density. These methods were based on interactive or automatic thresholding and segmentation. Some examples of such commercially available software are Cumulus [3], Volpara [4] and Quantra [5]. Mammographic Percent Density (MPD) is used in these software for assigning breast density categories. MPD is the area ratio of fibroglandular tissue to total breast tissue.

The state of mammography acquisition has changed considerably in the last few years. Instead of acquiring full-field 2D digital mammograms (FFDM), many clinics are offering 3D tomosynthesis acquisitions with reconstructed 2D images. These novel scanning techniques are favored as they have been shown to increase cancer detection and reduce false positive exams. These scans are becoming more popular compared to traditional 2D full-field digital mammography.

Early attempts at breast density estimation were comprised of handcrafted features applied to a very small dataset ($< 500$ images) [6]. In [7], the authors used Fuzzy C-means for segmenting pixels with similar tissue appearance followed by extraction of texture features of each segmented group, which are used for training different classifiers. The dataset was limited to 300 mediolateral oblique (MLO) views. Statistical approaches to breast density segmentation were used to segment internal parenchyma into either fatty or dense class [7]. The efficacy and generalizability of such feature-based methods have been questioned as neural networks and large datasets have taken the center stage. Automatic breast density classification using neural networks [8] has been the trend in the last decade. Researchers in [9] have used a convolutional neural network to estimate breast density. In [10], authors have used federated learning across five institutions to train a neural network to mitigate data scarcity and increase the generalizability of the model. The majority of papers using neural networks can be classified into either a transfer learning technique or a patch-based technique. In a transfer learning setup, a pre-trained model typically trained on a large dataset like ImageNet is repurposed for classifying breasts into different density categories. In contrast, the patch-based techniques use texture-based information for training deep neural networks.

In this paper, we present a deep learning based technique which uses transfer learning for breast density classification. The goal of this paper is to show that our accuracy increases once we are more inclusive of the imaging protocols. Most of the above mentioned methods use FFDM scans for training a breast density classifier. We showed that the models trained using FFDM scan perform poorly on images with different imaging protocols. We suggest that instead of restraining ourselves to a single acquisition protocol, we can combine all the images and train a model with comparable performance. One major advantage of training such a model is continuous learning where we can improve the model by adding more data as they are acquired at the imaging centers.

## 2. DATA

The data is collected from the Mayo Clinic Picture Archive and Communication Systems (PACS) from June to September 2021. The dataset comprised of a mix of Full Field Digital Mammography (FFDM), Hologic's 2D Intelligent view and C-view and 2D tomographic projections. Sample images for each acquisition type are shown in Figure 1. A regular mammography exam consists of four images. For each breast two images are acquired, one medio-lateral oblique view and one cranio-caudal view. Often times additional images like XCCM and XCCL (exaggerated cranio-caudal medial and lateral respectively are acquired. Table 1 shows the distribution across different acquisition protocols. The mammogram images are collected from 26,411 patients. The age-based demographics are shown in the table 2. The number of patients in training, validation and tests sets for each density group is shown in table 3.

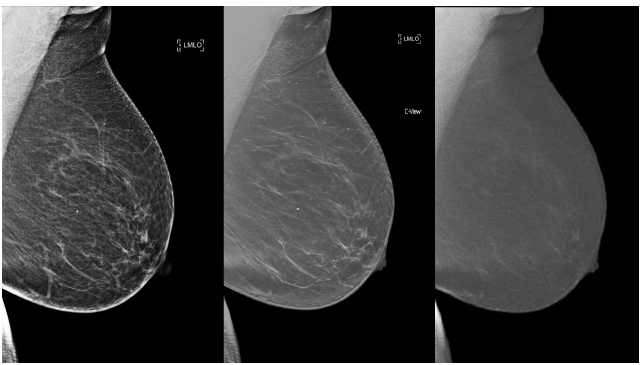

**Fig. 1**. Sample images for different acquisition types. Left-to-right: FFDM, C-View, Tomosynthesis Projection.

No patients in the aforementioned duration were excluded in the study. Figure 1 shows the difference in intensity profiles for FFDM, C-Views and Tomosynthesis projection.

## 3. METHODS

Transfer learning is a well established technique that is widely used when the data is limited. Although the mammogram dataset has comparatively large number of images as compared to other medical imaging data, it is still small compared to the general computer vision dataset like ImageNet [11], Microsoft COCO and Pascal VOC. We used the Inception-V3 [12] model with pre-trained weights (on ImageNet dataset) as the initial weights. For transfer learning, we replaced the last layers of the original Inception-V3 model with a fully connected layer containing 1024 nodes followed by four output nodes with softmax activation. All training was performed using Keras with TensorFlow-1.10. The initial learning rate was 0.001 on a stochastic gradient descent optimizer with batch size of 8; re-training was terminated after 50 epochs. Traditional augmentation routines (e.g., random rotation, horizontal and vertical flipping, random crops, and translation) were performed for each case. During the training/ validation process, algorithm performance (monitoring binary cross-entropy) on the validation set was observed per epoch with preservation of the model with highest accuracy to that point; if the validation accuracy increased in subsequent epochs, the model was updated. We trained four models based on different acquisition protocols. These models are 1. Tomosynthesis 2D projection, 2. FFDM images, 3. Hologic's 2D Synthetic images. and 4. A combined model trained based on all the three datasets.

**Table 1**. Data distribution across different acquisitions

| Acquisition | Number |
|---|---|
| C-View | 68,935 |
| FFDM | 51,727 |
| Tomosynthesis Projection | 15,384 |
| Intelligent 2D | 5,250 |
| XCCL | 3,199 |
| XCCL C-View | 2,272 |
| XCCL Tomosynthesis Projection | 1,752 |
| Latero medial | 609 |
| Latero-medial Tomosynthesis Projection | 427 |
| Medio-lateral | 327 |
| Medio-lateral C-View | 269 |
| XCCM | 159 |
| XCCM C-View | 124 |
| XCCL Intelligent 2D | 130 |
| **Total** | **151,164** |

**Table 2**. Age based demographics

| Age (years) | Number of patients |
|---|---|
| < 40 | 218 |
| 40-49 | 4,678 |
| 50-59 | 6,964 |
| 60-69 | 8,169 |
| ≥ 70 | 6,382 |
| Total | 26,411 |

**Table 3**. Number of patients in train, validation and test set in each breast density group (A, B, C, D).

| Group | Training | Validation | Testing |
|---|---|---|---|
| A | 1,847 | 614 | 618 |
| B | 6,946 | 2,312 | 2,316 |
| C | 5,862 | 1,952 | 1,954 |
| D | 1,193 | 396 | 401 |

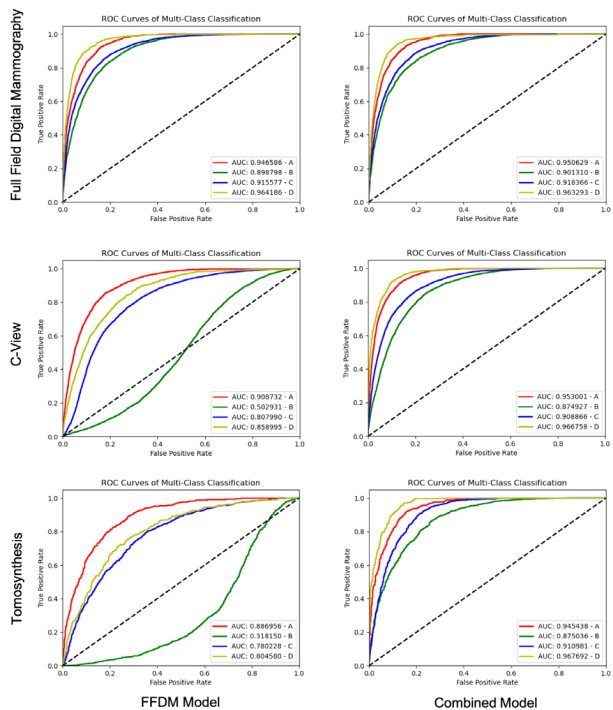

**Fig. 2**. AUC-ROC curves for testing set for different acquisition protocols. Top: Full Field Digital Mammography. Middle: Hologic's C-View. Bottom: 2D Tomosynthesis Projection.

## 4. RESULT

In this paper, we focus on real clinical data instead of restricting our models to full field digital mammography (FFDM) datasets. Our results are shown in Figure 2. The first column shows the performance of a FFDM model on FFDM, tomosynthesis and Intelligent 2D dataset. The second column shows the performance of a combined model on each of the above mention datasets. The first row shows that the FFDM model and combined models give comparable performance on the FFDM test set. The second and the third rows show the performance on FFDM model and the combined model on tomosynthesis data and Intelligent 2D datasets. We show an improved performance in all four categories of breast density measurement when a combined model is used. The increase in performance can be attributed to a more general model because of the increase in data size. As shown in table 1, the FFDM images comprise only one-third of the image population and thus are less useful for real world deployment and automatic reporting. The other reason for increased performance is natural data augmentation. At the Mayo Clinic, different image acquisitions are made for the same patient so different intensity profiles are available for the same underlying tissue as shown in Figure 1. This provides a natural intensity based data augmentation, which helps in increasing the model performance across all test sets.

## 5. CONCLUSIONS

In this paper, we showed that it is possible to train a neural network by mixing different acquisition protocols. The major contribution of this paper is to show that training a comparable neural network model for breast density classification is possible using the alternate low dose acquisition protocols in current clinical use. We have presented a real-world scenario in which a model is trained on clinical data as it is acquired. We showed that even if the model is trained on datasets containing images from a variety of protocols, it produces comparable and in some cases better results than a standard FFDM model. Another goal of this paper is to show that researchers should not shy away from mixing images from different protocols. In fact, this is key in producing a more generalizable neural network model which can be used for continuous learning as more and more data is collected. In the future, we would like to implement a Federated Learning model for further generalization.

## 6. REFERENCES

[1] Carl D'Orsi, L Bassett, S Feig, et al., "Breast imaging reporting and data system (BI-RADS)," *Breast imaging atlas*, 2018.

[2] John N Wolfe, "Risk for breast cancer development determined by mammographic parenchymal pattern," *Cancer*, vol. 37, no. 5, pp. 2486–2492, 1976.

[3] Jeffrey William Byng, NF Boyd, E Fishell, RA Jong, and Martin J Yaffe, "The quantitative analysis of mammographic densities," *Physics in Medicine & Biology*, vol. 39, no. 10, pp. 1629, 1994.

[4] M Jeffreys, J Harvey, and R Highnam, "Digital

mammography," 2010.

[5] Stefano Ciatto, Daniela Bernardi, Massimo Calabrese, Manuela Durando, Maria Adalgisa Gentilini, Giovanna Mariscotti, Francesco Monetti, Enrica Moriconi, Barbara Pesce, Antonella Roselli, et al., "A first evaluation of breast radiological density assessment by quantra software as compared to visual classification," *The Breast*, vol. 21, no. 4, pp. 503–506, 2012.

[6] Keir Bovis and Sameer Singh, "Classification of mammographic breast density using a combined classifier paradigm," in *4th international workshop on digital mammography*. Citeseer, 2002, pp. 177–180.

[7] Arnau Oliver, Jordi Freixenet, and Reyer Zwiggelaar, "Automatic classification of breast density," in *IEEE International Conference on Image Processing 2005*. IEEE, 2005, vol. 2, pp. II–1258.

[8] D Arefan, A Talebpour, N Ahmadinejhad, and A Kamali Asl, "Automatic breast density classification using neural network," *Journal of Instrumentation*, vol. 10, no. 12, pp. T12002, 2015.

[9] Nan Wu, Krzysztof J Geras, Yiqiu Shen, Jingyi Su, S Gene Kim, Eric Kim, Stacey Wolfson, Linda Moy, and Kyunghyun Cho, "Breast density classification with deep convolutional neural networks," in *2018 IEEE International Conference on Acoustics, Speech and Signal Processing (ICASSP)*. IEEE, 2018, pp. 6682–6686.

[10] Holger R Roth, Ken Chang, Praveer Singh, Nir Neumark, Wenqi Li, Vikash Gupta, Sharut Gupta, Liangqiong Qu, Alvin Ihsani, Bernardo C Bizzo, et al., "Federated learning for breast density classification: A real-world implementation," in *Domain Adaptation and Representation Transfer, and Distributed and Collaborative Learning*, pp. 181–191. Springer, 2020.

[11] Alex Krizhevsky, Ilya Sutskever, and Geoffrey E Hinton, "Imagenet classification with deep convolutional neural networks," *Advances in neural information processing systems*, vol. 25, pp. 1097–1105, 2012.

[12] Christian Szegedy, Vincent Vanhoucke, Sergey Ioffe, Jon Shlens, and Zbigniew Wojna, "Rethinking the inception architecture for computer vision," in *Proceedings of the IEEE conference on computer vision and pattern recognition*, 2016, pp. 2818–2826.